# Industry 4.0: Contributions of Holonic Manufacturing Control Architectures and Future Challenges

William Derigent*[1], Olivier Cardin**, Damien Trentesaux***

*Research Centre for Automatic Control of Nancy, CNRS UMR 7039, Campus Sciences, BP 70239, Vandoeuvre-lès-Nancy Cedex, 54506, France (e-mail: william.derigent@univ-lorraine.fr)

** LUNAM Université, Université de Nantes, LS2N UMR CNRS 6004
2 avenue du Pr Jean Rouxel, 44470 Carquefou, France (e-mail: olivier.cardin@ls2n.fr)

*** LAMIH UMR CNRS 8201, University of Valenciennes and Hainaut-Cambrésis, UVHC
Valenciennes, France (e-mail: damien.trentesaux@univ-valenciennes.fr)

Abstract: The flexibility claimed by the next generation production systems induces a deep modification of the behaviour and the core itself of the control systems. Over-connectivity and data management abilities targeted by Industry 4.0 paradigm enable the emergence of more flexible and reactive control systems, based on the cooperation of autonomous and connected entities in the decision-making process. From most relevant articles extracted from existing literature, a list of 10 key enablers for Industry 4.0 is first presented. During the last 20 years, the holonic paradigm has become a major paradigm of Intelligent Manufacturing Systems. After the presentation of the holonic paradigm and holon properties, this article highlights how historical and current holonic control architectures can partly fulfil I4.0 key enablers. The remaining unfulfilled key enablers are then the subject of an extensive discussion on the remaining research perspectives on holonic architectures needed to achieve a complete support of Industry4.0.

## 1 Introduction

The evolution of industrial systems to the so-called Industry 4.0 is mainly based on the development of highly connected resources throughout the whole process. This constant flow of information spread by and available for all the actors opens many opportunities to enhance the efficiency of the whole process, in the different phases of a product lifecycle. Direct benefits are expected in several domains among which:

- *Design*: Advanced digital tools enable disruptive product development with both physical and virtual prototyping for the design of smart products, for example with additive manufacturing or virtual reality (Nunes, Pereira, & Alves, 2017);
- *Logistics/supply*: information immediately transmitted to the whole supply chain enables a constant synchronization between stakeholders and an easier adaptation to unforeseen changes;
- *Manufacturing*: Using communicating objects as elements of the production line allows the on-line collection of information related to the production system. It is then possible to adjust the behavior of the shop-floor in real-time in case of abnormal conditions or changes (Mrugalska & Wyrwicka, 2017);
- *Maintenance and recycling*: These information could lead to new simulation-based monitoring and knowledge management abilities (Bokrantz, Skoogh, Berlin, & Stahre, 2017).

This horizontal transversality all along the product lifecycle is well expressed in the RAMI 4.0 model (*Reference Architecture Model for Industry 4.0*) proposed by (Zezulka, Marcon, Vesely, & Sajdl, 2016) and showed on figure 1. In this three-dimensional framework, the multi-scale aspects of the Industry 4.0 (e.g. from the workshop to the supply chain) as well as the different viewpoints on the benefits of the Industry 4 .0 are coupled with a dynamic vertical data integration (from assets to business process) more massive and intense than it is currently. It illustrates the numerous remaining challenges still to be addressed (Jose Barbosa, Leitao, Trentesaux, Colombo, & Karnouskos, 2017). The Industry 4.0 assumes a real-time adaptability of the whole company to external/internal changes (perturbations during production or supply, changes in customer-supplier relationships, and so on). To allow an efficient use of data available to manage these Industry 4.0-oriented processes, it becomes crucial to define enterprise information systems equipped with connected, interoperable, flexible and reactive control architectures.

---

[1] Corresponding Author

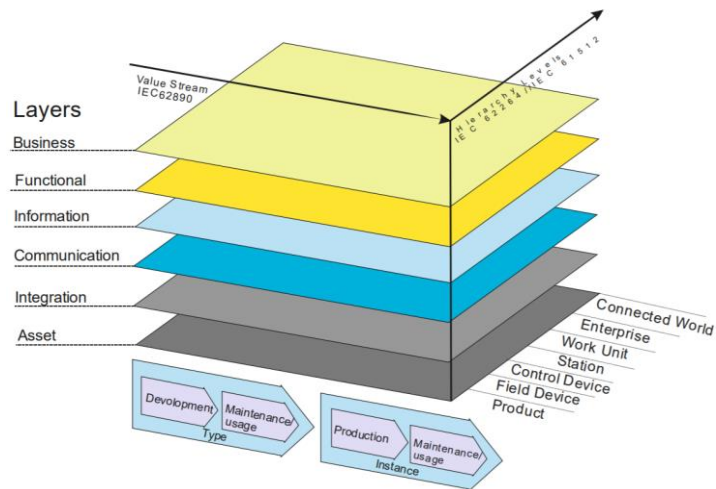

**Figure 1. Modèle RAMI4.0** (Zezulka et al., 2016)

In the manufacturing domain, many research works address the problem of manufacturing control. In order to tackle this problem, among possible approaches, one is to consider a manufacturing shop-floor as a system composed of autonomous entities cooperating together to achieve their own objectives and consequently the system goals. (A Koestler, 1967) postulates a set of underlying principles to explain the self-organizing tendencies of social and biological systems. He proposed the term *holon* to design the components of these systems. This term is a combination of the Greek word *holos*, meaning "whole", with the suffix *-on* meaning "part" (S Bussmann, 1998; Arthur Koestler, 1979). A holon consists of an information processing part combined most of the time with a physical processing part. Holons are recursive meaning that holons can form a recursive organisation. Moreover, a group of holons can cooperate to form a hierarchy referred to as *holarchy.* The informational part of a holarchy is referred to as a Holonic Control Architecture (HCA). During the last 20 years, HCAs have been widely studied and developed in the manufacturing domain (Morel, Pereira, & Nof, 2019). Several review papers (P Leitão, 2009; L Monostori, Váncza, & Kumara, 2006; Shen, Hao, Yoon, & Norrie, 2006; Shen & Norrie, 1999; Trentesaux, 2009) have presented HCAs and stressed their advantages and drawbacks. Those were especially exhibited by the various industrial applications of the paradigm that are described in literature, from early applications in automotive industry (Stefan Bussmann & Sieverding, 2001) to railway transportation systems (Le Mortellec, Clarhaut, Sallez, Berger, & Trentesaux, 2013) and radiopharmaceutical products (Borangiu, Ruaileanu, Oltean, & Silicsteanu, 2019).

The objective of this paper is not to provide an exhaustive survey of existing holonic architectures. This paper rather intends to give an overview on the evolution of HCAs over the last two decades and to explain how they can contribute to the dissemination of Industry 4.0 technologies in the manufacturing domain. It will also stress how these HCAs can evolve to better fit the Industry 4.0 needs and will suggest challenging research issues to address in the near future. The paper is thus structured as follow: the section 2 introduces the key enablers of the so-called Industry 4.0, which is a major paradigm shift in Manufacturing (Almada-Lobo, 2016). Section 3 will briefly define the fundamentals of HCAs. The different proposed architectures are detailed in section 4 followed by a comparison of the properties of these architectures in regards of the key enablers of Industry 4.0. Based on this last comparison, section 5 will state the remaining challenges needed to completely fit the Industry 4.0 prerequisites and described the remaining research perspectives to develop new kinds of HCAs for Industry 4.0.

## 2  Key enabling characteristics of Industry 4.0 in manufacturing

Without the ambition to define what is Industry 4.0, our work needs a set of key enablers of Industry 4.0 in the manufacturing domain to study and position existing holonic control architectures in regards. This section aims at elaborating such a set.

The term Industry 4.0 was first introduced in 2011 by (Kagermann, Lukas, & Wahlster, 2011). Today, digitalization, interconnectedness and new manufacturing technologies are hoped to bring new business models, sustainable and efficient use of limited resources and cost-effective production of highly customized products as stated by (Schuh, Anderl, Gausemeier, ten Hompel, & Wahlster, 2017). All the related research and technological developments are grouped under the term "Industry 4.0" defined as "real-time, high data volume, multilateral communication and interconnectedness between cyber-physical systems and people". (Hermann, Pentek, & Otto, 2016) consider Industry 4.0 as "a collective term for technologies and concepts of value chain organization". To achieve the Industry 4.0 objectives, future Industry4.0-compliant systems should respect several requirements. (Liao, Deschamps, Loures, & Ramos, 2017) propose an extensive study of papers related to the Industry 4.0. A part of this study focuses on the research initiatives trying to list the different requirements or enabling features an Industry4.0-compliant system should respect or implement. The authors select a panel of 176 related papers and identify a short list of most cited papers cf. Figure 2 (Berger, 2014; Brettel, Friederichsen, Keller, & Rosenberg, 2014; Hermann et al., 2016; Kagermann,

Helbig, Hellinger, & Wahlster, 2013; Schuh et al., 2017; Vogel-Heuser & Hess, 2016). From an analysis of the positioning of those articles, a comprehensive list of prerequisites for Industry 4.0 compliant control systems is synthesized.

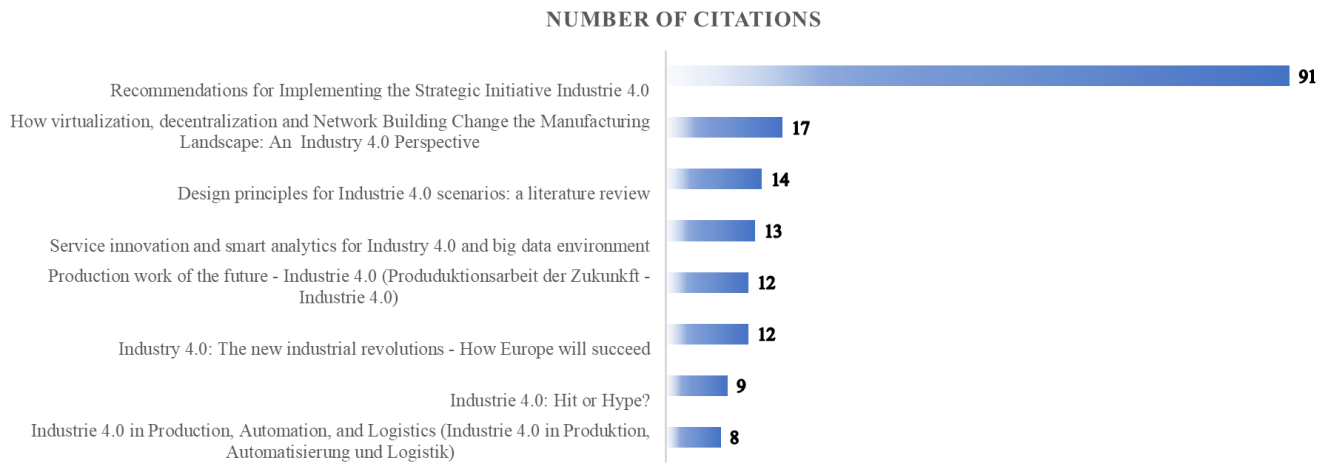

**Figure 2. Frequency of references (more than 5 times), from** (Liao et al., 2017)

Acatech, the German National Academy of Science and Engineering, characterizes the Industry 4.0 as "a new level of socio-technical interaction between all actors and resources involved in manufacturing, with networks of manufacturing resources that are autonomous, capable of controlling themselves in response to different situations, self-configuring, knowledge-based, sensor-equipped and spatially dispersed" (Kagermann et al., 2013). In these systems safety and security are crucial to protect humans and data privacy. Services must be available online through digital market places.

(Brettel et al., 2014) consider that Industry 4.0 is possible with systems composed of distributed self-controlling entities capable to monitor, acquire and process large amount of product/machine data. Products must be able to communicate with their environment and influence the arrangement of the manufacturing system. In this vision, manufacturing systems should have the abilities to form collaborative networks capable of exchanging information in a standardized form. Processes must be virtualized to enable the collective setup of simulation chains between manufacturers, fed by real-time data.

(Berger, 2014) sees the Industry 4.0 as smart robots and machines with smart sensored human-machine interfaces, generating huge amount of data that need to be secure. Innovative methods to handle Big Data will create new ways to leverage information. Moreover, the future manufacturing systems are sustainable, energy efficient and equipped with decentralized small-size plants to produce their own power. Virtual plants and products will be used with simulation tools to prepare the physical production.

(Hermann et al., 2016) consider 6 design principles to implement Industry 4.0, which are Interoperability, Virtualization, Decentralization, Real-Time Capability, Service Orientation and Modularity.

(Vogel-Heuser & Hess, 2016) consider that Industry 4.0 implies self-organization, autonomy, adaptability, big data analytics, data integration, secure communication and connectivity.

Another Acatech study (Schuh et al., 2017) identify some elements needed for the Industry 4.0: Real-time capability, Big Data Analysis, Decision Support Systems with Automated Decision Making, Vertical and Horizontal Process/Systems Integration and Cyber-Physical Systems.

The prerequisites enounced above vary between authors as they do not have the same point of view and do not adopt the same level of granularity. Indeed, these key enablers have inner relationships and must not be seen as independent characteristics. For example, connectivity is a mandatory element for all the other key enablers, big data analysis can improve other key enablers as adaptability. They however form a consistent vision of what should be the key enablers for Industry 4.0, i.e. the main characteristics that should be expected from such control systems. In table 1, those key enablers are listed, together with a short definition and their positioning towards the articles listed above. Those 10 key enablers are considered as the basis of the analysis grid for the rest of this article.

Section 3 introduces some fundamentals in holonic control architectures. Section 4 describes the different kinds of evolution of HCAs that occured along the last 20 years. Each part of the fourth section contains, when relevant, a table referencing the different related research works with regard to the different key enablers of Industry 4.0.

## 3   Fundamentals in holonic control architectures

A Holonic Control Architecture (HCA) is an architecture composed of holons, called holarchy (S Bussmann, 1998; A Koestler, 1967; Arthur Koestler, 1979). A holon is a communicating decisional entity (with inputs and outputs) composed of a set of sub-level holons and, at the same time, part of a wider organization composed of higher-level holons (recursion phenomenon also called Janus effect(Arthur Koestler, 1979)). It is important to note that a holon is also composed of a physical part associated to a digital one (that can be modeled as a digital agent, avatar, digital twin, …) and finally, holons, which are merged into a holarchy (here, a HCA), are able to decide according to a certain degree of autonomy within this holarchy. As introduced in the holon general architecture of (Christensen, 1994), holons have an **holon-holon interface** but must also contain a **human-holon interface** to enable the communication between a human and a holon.

**Table 1. Industry4.0 key enablers**

| Key enablers for Industry 4.0 manufacturing systems as a prerequisite | (Schuh et al., 2017) | (Brettel et al., 2014) | (Berger, 2014) | (Hermann et al., 2016) | (Vogel-Heuser & Hess, 2016) | Definitions |
|---|---|---|---|---|---|---|
| Sustainability | | | X | | | Industry 4.0 Manufacturing Systems should be developed with socio-ecological objectives, respecting humans as much as environment. |
| Secure communication/ Cyber-Resilience | X | X | X | | X | Industry 4.0 Manufacturing Systems should rest to the numerous threats for attacks or intrusions, especially in the control system (Yampolskiy, Horváth, Koutsoukos, Xue, & Sztipanovits, 2015) |
| Real-Time Capabilities | X | X | X | X | X | Industry 4.0 Manufacturing Systems should monitor and react in real-time to unforeseen events. |
| Process Virtualization | X | X | X | X | | Industry 4.0 Manufacturing Systems will comprise virtual plant models and simulation models, directly linked to sensors used to monitor processes. These virtual models can be used for immediate control or forecasting. |
| Service Orientation | X | | | X | X | Industry 4.0 Manufacturing Systems should offer services via the Internet based on a service-oriented reference architecture. |
| Interoperability | X | X | X | X | X | Industry 4.0 Manufacturing Systems should include the necessary interoperability features required to communicate and exchange data/information/knowledge with other systems. |
| Adaptability | X | X | | | X | Industry 4.0 Manufacturing Systems should adapt their decisions to disturbances. |
| Big Data Analysis | X | X | X | X | X | Industry 4.0 Manufacturing Systems should learn (i.e generate new knowledge) from the past events. |
| Autonomous and decentralized Decision Support Systems | X | X | X | X | X | Industry 4.0 Manufacturing Systems will be composed of autonomous and distributed manufacturing entities. To control such systems, distributed decision support systems are needed. |
| Connectivity | X | X | X | X | X | Industry 4.0 Manufacturing Systems should be easily connected through any communication network. |

As stated in (Giret & Botti, 2004), HCAs exhibit some generic properties presented and discussed hereinafter.

**Decision making** is one fundamental property in HCAs. Researches on the topic of decision making by artificial entities have been led for several decades (Pomerol, 2013). In the industrial management community, deciding is the activity of reducing a set of possibilities. A close notion is the concept of "degree of freedom". From this point of view, classical decision-making activities can be derived: choice (set reduced to a singleton), ranking (integrating order in the set), etc. both in discrete (set of resources) and continuous (rotation speed, energy consumption) domains. In HCAs, given the importance of physical aspects, deciding is an activity merged into a wider process that can be called a decisional process. Extending the basic ideas of Simon, a decisional process is composed of several activities: monitoring, triggering, design of possible decisions, a-priori evaluation of decisions, decision, application and a-posteriori evaluation of decisions (Simon, 1996).

The decision process (the holon behavior) can be **reactive** (reflex as an automated response to a stimulus) or **pro-active** (goal-oriented response). A holon implementing a reactive behavior perceives its environment's stimuli and responds using pre-programmed behaviors. A pro-active holon is not only a stimulus-reacting entity but can also act to favor its own goals. Whatever happens, holons are **rational** in the sense that they always try to choose a (local and/or global) optimal decision. To take a good decision, holon can use **simulation or optimization models**.

Given the **recursive aspect of a holarchy**, this decision process is also recursive and can be implemented into layers of holons. For example, the triggering activity for a quality control holon can be decomposed into a decision process handled by lower level holons aiming to decide, through a learning strategy, the best triggering level to avoid over-reaction if too low or to avoid loss of customer if too high.

A holon evolves in an environment in constant evolution, hardly predictable. The pre-defined goals of a holon may not remain valid (or optimal) if a change occurs in the holon environment. **Adaptability** is then needed for the holon to modify its goals depending on its knowledge of the environment. This also means that a holon should learn for its experience and environment to be able to correctly adapt its behavior. Here, **learning** means generating new knowledge from data acquired on the manufacturing system.

**Autonomy** is another fundamental property. It is defined as the degree of freedom of each holon regarding its decision capacity, whatever the holon level (Giret & Botti, 2004). It can also be associated to a set of constraints on a search space when using optimization tools. The level of autonomy can be set during the design phase by the designer himself, but it can also be adjusted by a higher level holon with application to a lower level during the exploitation phase. For example, a supervisor holon decides to restrict the set of possible resource-holons to be chosen by lower-level product holons because of a maintenance operation to come on one of these resource-holons. Such decisional autonomy must be accompanied by an operational autonomy enabling the holon to apply its decisions and monitor the execution of its plans, with the ability to take corrective actions again perturbations.

**Cooperation** among holons enables to restrict or enlarge the autonomy of holons. For example, a direct peer-to-peer negotiation protocol (Smith, 1980) or an indirect through the environment use of pheromone-based communication (Pannequin, 2007; P. Valckenaers & Van Brussel, 2005) can lead holons to improve the quality of their decisions through the forbidding of search spaces to avoid local optima during a dynamic task allocation process. This property also includes the possibility to cooperate with a human operator.

At the level of the HCA, one important aspect is relevant to **openness**, that is the ability to integrate, suppress holons with a minimal external intervention ("plug and play" concept) as well the ability to modify the dependencies or the hierarchical relationships among holons also with a minimal external intervention (McFarlane & Bussmann, 2013).

Because of these characteristics for a holon and an HCA, it is expected that emerging behaviors occur (Maja J Mataric, 1993; C. Pach, Bekrar, Zbib, Sallez, & Trentesaux, 2012). From our perspective, an **emerging behavior** is the observation of a property at a higher level of a HCA that has not been explicitly integrated (programmed) into holons composing this HCA. For example, using attractive/repulsive potential field algorithms may lead product-holons to naturally avoid a resource-holon under breakdown and to select it again after recovery without explicitly coding this behavior in the different product-holons composing the HCA. Emerging behaviors are sometimes positive regarding the whole system performance but can also lead to deadlocks or inefficient situations.

The assumption leading this research paper is that HCAs, to a certain extent, can fulfill the introduced key enablers of Industry 4.0 manufacturing systems. The question is thus about the consistency of introduced HCA properties with regard to those key enablers. For example, the key enabler "Autonomous and Decentralised Decision Support Systems" corresponds to "Decision-Making" and "Autonomy" properties, while the key enabler "Adaptability" corresponds to the "Adaptability' property of the HCAs, cf. table 2. From our point of view, several key enablers posed by the Industry 4.0 theoretically fit with HCAs properties. However, even if the holon paradigm seems to be consistent with the one of Industry 4.0, research works led on HCAs may only cover some of the required holon properties, and consequently, may cover only parts of the prerequisite key enablers of the Industry 4.0 manufacturing system.

The aim of the next section is thus to analyse the state-of-the-art contributions in the field of HCAs applied to manufacturing systems regarding these key enablers. These contributions are grouped into 6 categories (Historical, Dynamic, Data-oriented, Product-Centric, Web-oriented and Digital-twin-based HCAs). For that purpose, the title of each category recalls the different Industry 4.0 key enablers concerned by the studied category.

# 4 Contributions of Holonic Control Architectures to Industry 4.0 manufacturing systems

Implementing a control system on a set of autonomous communicating entities is a difficult research problem. Many research works study HCAs and propose architectures with some specific properties. Among them, some are the "pioneering" ones that could be referred to as "Historical Architectures": indeed, they are often the ones used as a basis to define new architectures. Some other architectures focus on the "adaptability" property of the holonic systems and include reorganization mechanisms at the structure level. They are called "Dynamic Architectures". Some try to study "learning/aggregation" mechanisms and are defined as "Data-oriented Architectures". Some derive historical architectures to specifically manage data related to the product holons and are called "Product-Centric Architectures".vThe "Web-Oriented Architectures" have been defined to be compatible with the cloud technologies and more recently, the "Digital-Twin based Architectures" integrate the notion of digital twin.

**Table 2. Industry4.0 manufacturing systems key enablers and corresponding holonic properties**

| Key enablers for Industry 4.0 manufacturing systems as a prerequisite | Corresponding Holon Properties |
|---|---|
| Sustainability | Energy efficiency can be defined as a goal for a holon. Human/Machine Interfaces are a core component of the holon. |
| Secure communication/Cyber-Resilience | No clear properties. |
| Real-Time Capabilities | Reactivity, Adaptability |
| Process Virtualization | Simulation and optimization models used during decision |
| Service Orientation | No associated property. However, the distributed and autonomous nature of holons ease the definition of services. |
| Interoperability/Integration | Cooperation, Openness |
| Adaptability | Adaptability |
| Big Data Analysis | Learning/Aggregation |
| Autonomous and decentralized Decision Support Systems | Reactivity, Pro-Activity, Autonomy |
| Connectivity | Cooperation |

## 4.1 Historical Architectures: towards connected, autonomous, decentralized decision support systems

Historical architectures are a category regrouping all the "pioneering" reference architectures proposed at the early stages of HCAs. These first architectures implement some properties of the holonic paradigm such as cooperation, autonomy and decision-making (reactive and pro-active).

(Van Brussel, Wyns, Valckenaers, Bongaerts, & Peeters, 1998) described the first reference architecture issued from an IMS Project called Holonic Manufacturing Systems in 1996. Its acronym is PROSA standing for Product-Resource-Order-Staff Architecture (figure 3) as the 4 holons composing the architecture:

- Product holons, managing the production knowledge;
- Resource holons, managing the process knowledge;
- Order holons, managing the process execution;
- Staff holons, not represented in the figure, acting as a global advisor for the whole architecture.

This architecture is the most referenced one (over 900 times early 2019 in Scopus Database[2]) in the literature. It is often the basis of subsequent architectures as being the most generic one.

This architecture was the first of a series, developed for different domains (namely manufacturing, logistics, maintenance, etc.). For example, *PROSIS* (Product, Resource, Order, Simulation for Isoarchy Structure) was designed to offer a different organization paradigm using the concept of isoarchy (Pujo, Broissin, & Ounnar, 2009). In an isoarchy, there are no subordination hierarchical links between holons. At a same decision level, the different decision-making entities are equal in the decision-making mechanism. In the PROSIS architecture, the Staff Holon, useless in such isoarchic context, has been replaced by the

---

[2] https://www.scopus.com/authid/detail.uri?authorId=7004590058

Simulation Holon able to simulate the evolution of the manufacturing system from its current status, computed via the active listening and analysis of interactions between all other holons.

Implementing the architectures in a control system perspective is a question that needs to be investigated, and several architectures were derived from PROSA in order to benefit from the theories and tools developed in the multi-agent field. Among them, *HCBA* (Holonic Component Based Architecture) (Chirn & McFarlane, 2000) is the first architecture based on a fusion of different concepts originating from component-based development, multi-agent system (MAS) and HMS. The purpose of such fusion is to develop a highly decentralized architecture, built from autonomous and modular cooperative and intelligent components, able to manage rapidly the different changes, by focusing on the system reconfigurability.

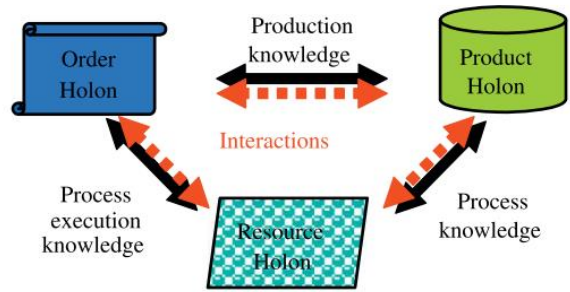

**Figure 3. PROSA simplified architecture**

HCBA is composed of 2 types of components in the production system, Resource Component and Product Component. The Resource Component is composed of a physical part and a virtual control part. Resources manage the operation scheduling, while looking for the optimization of their use. The Product Component is composed of a physical part and an informational one. Its physical part can represent materials, parts, pallets and so on. Moreover, the informational part manages the production program, including the routing control, the process control, decision-making and production information. The informational part is composed of virtual agents with specific roles. Each Product Component is referring to a Product Coordinator creating WIP agents (Work in Process). Both monitor the completion of orders, but at different levels. the Product Coordinator ensures the production monitoring of a lot whereas WIP agents are in charge of the production monitoring of an individual part. As a result, WIP agents negotiate with the resource community to define the part processing in the shop-floor. These negotiations are done within an objective set by the Product Coordinator (figure 4).

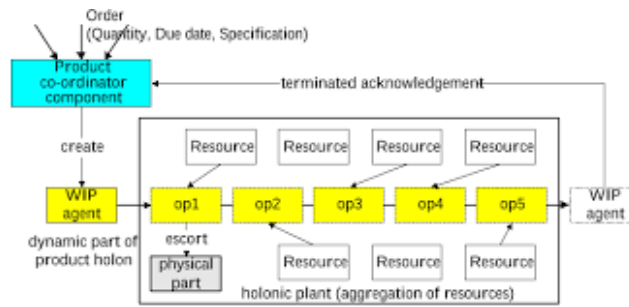

**Figure 4. Structure of the HCBA architecture**

In the same way, *Delegate MAS* is an architectural pattern that allows an agent to delegate a responsibility to a swarm of lightweight agents to support this agent in fulfilling its functions. The issuing agent can delegate multiple responsibilities, each of them applying the delegate MAS pattern. The agent may use a combination of delegate multi-agent systems to handle a single responsibility. The delegate MAS may also provide services to other agents. Delegate MAS is a more generic description of an approach inspired by ant food foraging behavior. The delegate MAS pattern consists of three elements: the agent, the ant and the environment (figure 5).

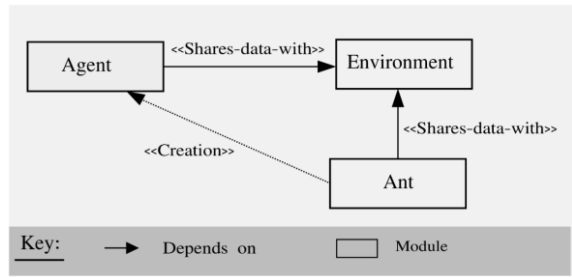

**Figure 5. Delegate MAS architectural view** (Verstraete et al., 2008)

*ADACOR* (ADAptive holonic COntrol aRchitecture) is a holonic reference for the distributed manufacturing system, proposed by (Paulo Leitão & Restivo, 2006). This architecture has a decentralized control architecture but also considers centralization in order to tend to a global optimisation of the system. Holons are belonging to the following classes: Product Holons (ProdH), Task Holons (TH), Operation Holons (OpH) and Supervisor Holons (SupH), interconnected via the scheme depicted figure 6.

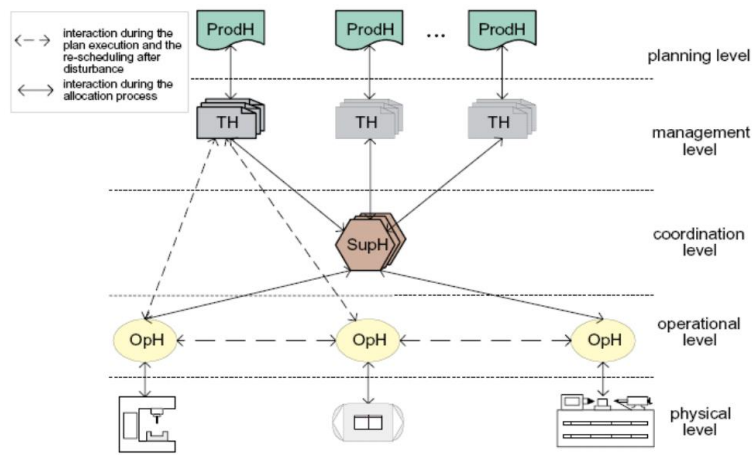

**Figure 6. ADACOR holon repartition** (Paulo Leitão & Restivo, 2006)

SupH are based on biological systems and are different from SH of PROSA. In normal execution, the ADACOR architecture maintains the production system in stationary state, where holons are organised in a hierarchical structure, the OHs following the optimised schedules proposed by the SupHs, for the THs. However, when a manufacturing problem occurs (delay, machine failure, …), the global system enters in transient state, characterised by the re-organisation of the holons required to react to the disturbance. To do so, ADACOR uses a pheromone-like spreading mechanism to distribute global information. Thanks to this one, ADACOR introduces the possibility to change dynamically the holarchy between the stationary state and the transient one.

These historical architectures form the foundation of the HCA domain. They have been used and refined in different research directions. In the following sections 4.2 to 4.5, these refinements and improvements are detailed. Each section contains a table summarizing the contributions of the different section-related works to the industry 4.0 key enablers.

### 4.2 Dynamic architectures: towards adaptability and better real-time capabilities

HCAs naturally seek for reactivity since they have been designed in that sense, to overcome the main drawback of centralized architectures. Hence, all the previous historical architectures can react quickly to disturbances. However, the global behavior of the system is constant and does not change. As a result, designing Dynamic control architectures is probably one of the most promising current trends in literature in HCAs (Cardin et al., 2017). It postulates that the behavior of the system can be changed dynamically in order to adapt to the changes of the environment and thus reduce the transient states and the associated loss of performance. There have been several dynamic architectures proposed in the literature, all of them being characterized by a "switching mechanism" enabling them to switch from a holonic architecture to another one. (Jimenez, Bekrar, Zambrano-Rey, Trentesaux, & Leitão, 2017) made a survey of the different kind of switching mechanisms and their use in dynamic holonic architectures.

In our review, 4 of these dynamic architectures have been highlighted. ORCA (Cyrille Pach, Berger, Bonte, & Trentesaux, 2014) was one of the first dynamic architectures that was formalized in literature (figure 7). In ORCA, a global optimizer Holon controls at a lower level local optimizer Holons. A switching mechanism, allowing local Holon to decide, occurs if a perturbation happens and forbids the application of a predefined schedule. This switch enables the Holonic architecture to adapt to perturbations but there is no switching back mechanism. In (Borangiu, Rəileanu, Berger, & Trentesaux, 2015), a mechanism is proposed to switch between a centralized and a decentralized holonic architecture in the presence of perturbations to ensure as long as possible both global optimisation and agility to changes in batch orders during manufacturing. This switching is bidirectional (figure 8).

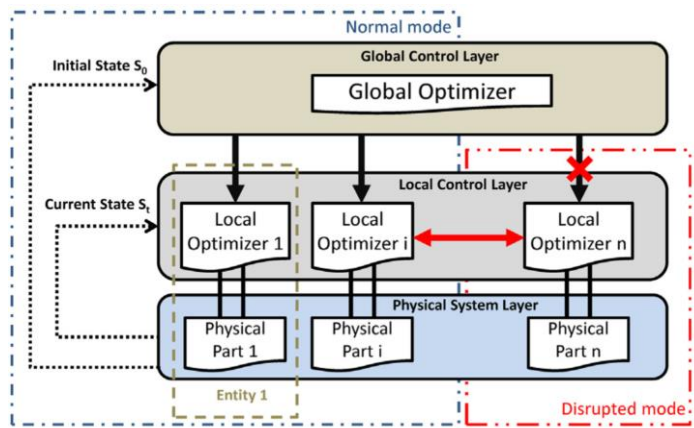

**Figure 7. ORCA global organization** (Cyrille Pach et al., 2014)

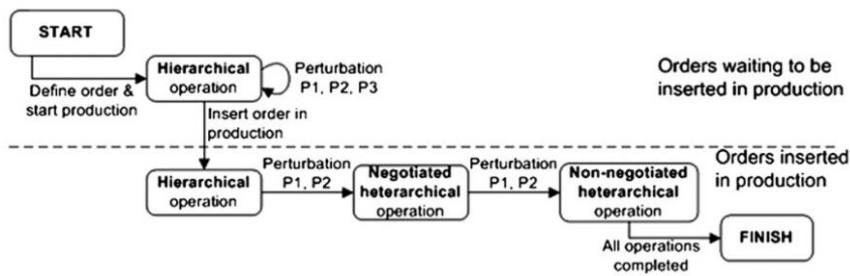

**Figure 8. An example of a switching process among holonic architectures** (Borangiu et al., 2015)

An evolution of ADACOR mechanism has also been presented in (José Barbosa, Leitão, Adam, & Trentesaux, 2015) as ADACOR². The objective is to let the system evolve dynamically through configurations discovered online, and not only between a stationary and one transient state (figure 9). The rest of the architecture is nevertheless quite similar to ADACOR.

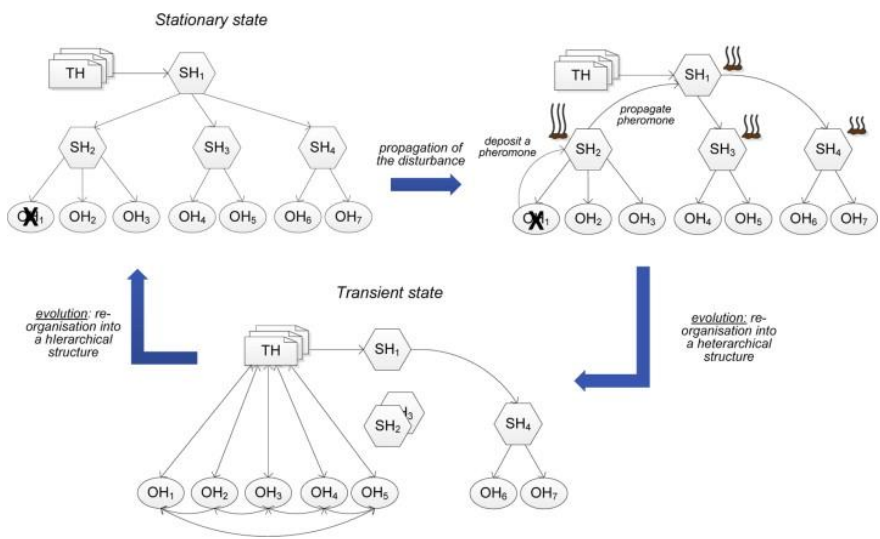

**Figure 9. ADACOR² evolution** (José Barbosa et al., 2015)

The last architecture in date is denoted as POLLUX (Jimenez et al., 2017). The main novelty is focused on the adaptation mechanism of the architecture, using governance parameters that enlarge or constrain the behaviour of the low level holons regarding the disturbances observed by the higher level. The simulation of the consequence on performances of several switching options ("what-if" scenarios) is proposed since in POLLUX, the number of possible switches at a given time is high compared to ORCA for example. Each decided switch is thus justified by an increase in the performances compared to other switching possibility (Figure 10).

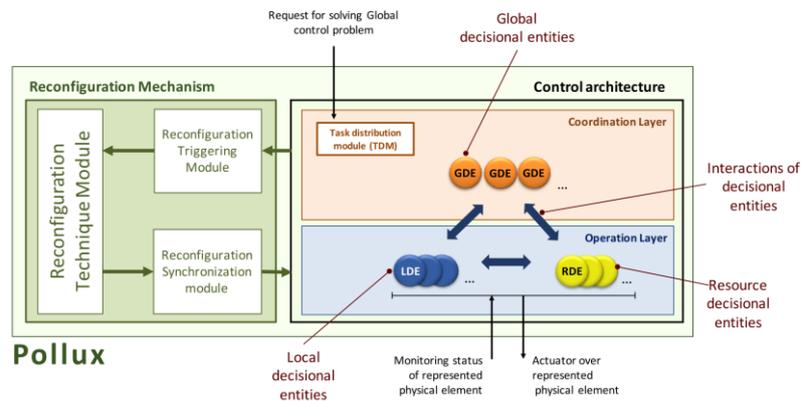

**Figure 10. POLLUX Mechanism** (Jimenez et al., 2017)

The advantages of these kind of architectures come from their ability to better cope with uncertainties in real-time: the architecture is adjusted according to real-time events. Meanwhile, these architectures suffer from two main drawback related to their possible nervousness (switching too many times, too fast) and their possible temporal myopia (a switching done that is in fact not necessary). The cost of switching is thus not really investigated.

**Table 3. Dynamic architectures vs industry4.0 manufacturing systems key enablers**

| Key enablers for Industry 4.0 manufacturing systems as a prerequisite | ORCA (Cyrille Pach et al., 2014) | ADACOR[2] (José Barbosa et al., 2015) | POLLUX (Jimenez et al., 2017) | (Borangiu et al., 2015) | Comments |
|---|---|---|---|---|---|
| Sustainability | | | | | Sustainability was not directly addressed, but some tries to handle energy consumption was initiated |
| Secure communication/ Cyber-Resilience | | | | | None of the dynamic architectures addresses this key enabler |
| Real-Time Capabilities | X | X | X | X | This capability is increased by for dynamic architectures |
| Process Virtualization | | | X | | Testing different switching strategies implied the virtualization of processes to simulate them, only POLLUX proposes it |
| Service Orientation | | | | | None of the dynamic architectures addresses this key enabler |
| Interoperability | | | | | None of the dynamic architectures addresses this key enabler |
| Adaptability | X | X | X | X | This is the flagship ability of dynamic architectures |
| Big Data Analysis | | | | | None of the dynamic architectures addresses this key enabler |
| Autonomous and decentralized Decision Support Systems | X | X | X | X | All the dynamic architectures are highly reactive and integrate holons whose autonomy is controlled |
| Connectivity | X | X | X | X | Cooperation is a natural mechanism for dynamic architectures |

## 4.3 Data-oriented Architectures: towards Big Data Analysis

In recent years, due to the advent of Big Data, the necessity to control data streams conveyed through the architecture emerged. SURFER architecture (Le Mortellec et al., 2013) is not specifically designed for manufacturing but is designed from the historical research in that field led by the team that designed it. It represents an interesting adaptation of control systems for maintenance and monitoring of highly complex systems (namely trains in this case). The generic holonic architectural model proposes for the diagnosis is shown in figure 11. This model is composed of recursive diagnosis structures, including sub-systems and their associated diagnosis methods. Each diagnosed system is composed of a control part and a controlled part, which operate in a context. The control part executes an algorithm to control the controlled part and, in return, the controlled part adopts an expected behaviour. At the lowest level of the holonic structure, the controlled part is typically composed of physical elements (e.g., sensors, switches and actuators) that are linked to mechanical and electrical constraints.

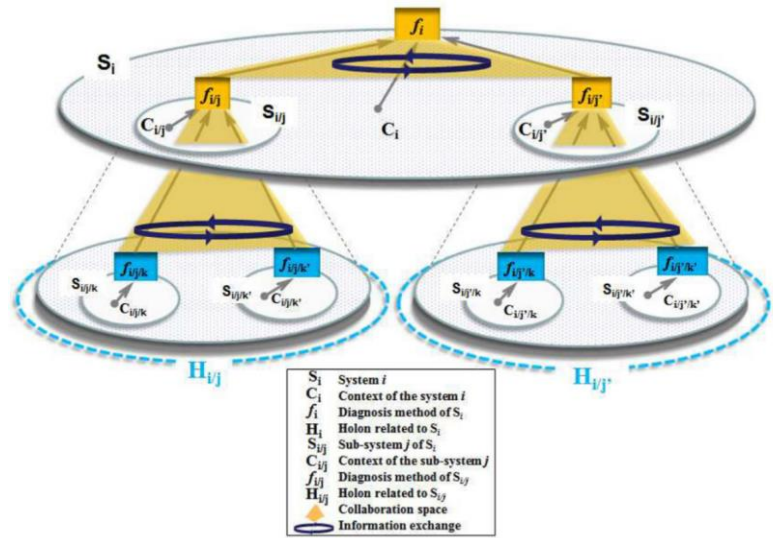

**Figure 11. Basics of the SURFER architecture**

This architecture is intended to handle a huge amount of data, as its direct application considers the monitoring of complex systems equipped with hundreds of sensors. In a traditional approach, the Big Data streams would have been gathered in a single place in order to characterize and understand the data and the relationship between them. In the SURFER architecture, the aggregation of holons makes it possible to create an architecture where the data are handled locally and propagated to the rest of the architecture in a higher semantic level. To do so, each holon can gather and understand the data from the lower level. These data are already treated and are only transmitted in a comprehensive way. With this mechanism, each layer of the architecture is providing some added value to the semantic of the data in order to reduce as much as possible the explosion of data to be treated by a single level (Trentesaux & Branger, 2018). Data are then sent when they are relevant.

## 4.4 Product-centric Holonic Control Architectures: towards product data interoperability

As stated before, Interoperability is the capacity of a system to communicate with another. Three levels of interoperability are commonly identified : technical, syntactic and semantic interoperability (EIF, 2004). The technical interoperability is related to physical connections of systems between them. The syntactic interoperability is related to communication and refers to the packaging and transmission mechanisms for data. The semantic interoperability stage is the last one and refers to the exchange data with an unambiguous and shared meaning.

Holonic Control Architectures have been used in works related to data management (Meyer, Främling, & Holmström, 2009) and dealing with interoperability. These works focus mainly on the concept of product holon (also named "intelligent product"). Architectures like EPC Global Network, Dialog or WWAI (Ranasinghe et al., 2010) or the later standard O-MI (The Open Group, 2014) consider that the physical part of the product holon is equipped with a tag (RFID, barcode, wireless sensor node) containing a product identifier. Thanks to this identifier, it is then possible to access to the associated informational part of the holon, often represented by an agent. Once reached, it is then possible to get/modify information from the agent related to the product instance, thus making possible to trace product evolution all along its lifecycle. As a result, all these works are regrouped under the term "Product Lifecycle Information Management". These approaches propose to any actor of the product lifecycle a certain number of mechanisms to remotely retrieve and update the product data, which may be interpreted as a syntactic product-centric interoperability.

Moreover, in these approaches, the product holon is considered as a data repository. Some works as (Ouertani, Baïna, Gzara, & Morel, 2011; Panetto, Dassisti, & Tursi, 2012; Terzi, Panetto, Morel, & Garetti, 2007) focus on the product data models while considering the product as an information mediation system in the sense of (Bénaben et al., 2008). The product is then capable to exchange data with all the other information systems it encounters during its lifecycle. Some works also propose product reference models (OntoPDM (Tursi, Panetto, Morel, & Dassisti, 2009), PRONTO (Giménez et al., 2008; Vegetti, Leone et Henning, 2011), Semantic Object Model (SOM) (Matsokis & Kiritsis, 2010)). These works, advocating that the holonic paradigm is a support to semantic interoperability, are summarized in table 4.

**Table 4. Product-centric architectures vs industry4.0 manufacturing systems key enablers**

| Key enablers for Industry 4.0 manufacturing systems as a prerequisite | | WWAI, EPC Global Network, DIALOG (Ranasinghe et al., 2010) | PRONTO, ONTO-PDM, SOM (Matsokis & Kiritsis, 2010; Panetto et al., 2012; Vegetti, Leone, & Henning, 2011) | Comments |
|---|---|---|---|---|
| Sustainability | | | | None of these architectures addresses this key enabler |
| Secure communication/ Cyber-Resilience | | | | None of these architectures addresses this key enabler |
| Real-Time Capabilities | | X | | These architectures support real-time data requests |
| Process Virtualization | | | | None of these architectures addresses this key enabler |
| Service Orientation | | X | | EPC Global Network and DIALOG implements a webservice-oriented data retrieval. |
| Interoperability | Technical | | | None of these architectures addresses this key enabler |
| | Syntactic | X | | These architectures allow to set up syntactic interoperability among different actors of the product lifecycle. They are product-centric. |
| | Semantic | | X | Works propose reference data models based on the holonic paradigm. |
| Adaptability | | | | None of these architectures addresses this key enabler |
| Big Data Analysis | | | | None of these architectures addresses this key enabler |
| Autonomous and decentralized Decision Support Systems | | | | None of these architectures addresses this key enabler |
| Connectivity | | X | X | Cooperation between product lifecycle actors is fostered by these architectures |

### 4.5 Web-oriented Architectures: towards Service-orientation and secure communication

One of the trends in developing HCAs is the evolution of cloud-based technologies. In this global trend, web-oriented HCAs integrate some of the web-based standards, such as web services. For example, SoHMS architecture (Gamboa Quintanilla, Cardin, L'Anton, & Castagna, 2016) is mainly based on principles and concepts introduced by PROSA, combining some interaction concepts from HCBA and ADACOR. This proposal uses the Product (PH), Resource (RH) and Order Holons (OH) from PROSA, and the concept of Directory Facilitator (DF) from multi-agent systems. Even if basic concepts remain close to the original ones, their behavior was adapted for making the services (namely Manufacturing Services – Mservices) the main element of interaction, oriented towards planning and scheduling activities. Figure 12 shows a class diagram of the architecture detailing the relations and data exchanges between actors. Because of the service perspective, a new element added to the architecture is the SIL (Service Interface Layer), used as an interface between service descriptions and their implementation methods at shop-floor. Due to its individual and proprietary characteristic, each resource possesses a SIL instance, containing all the information on the way to implement a service on the lowest level.

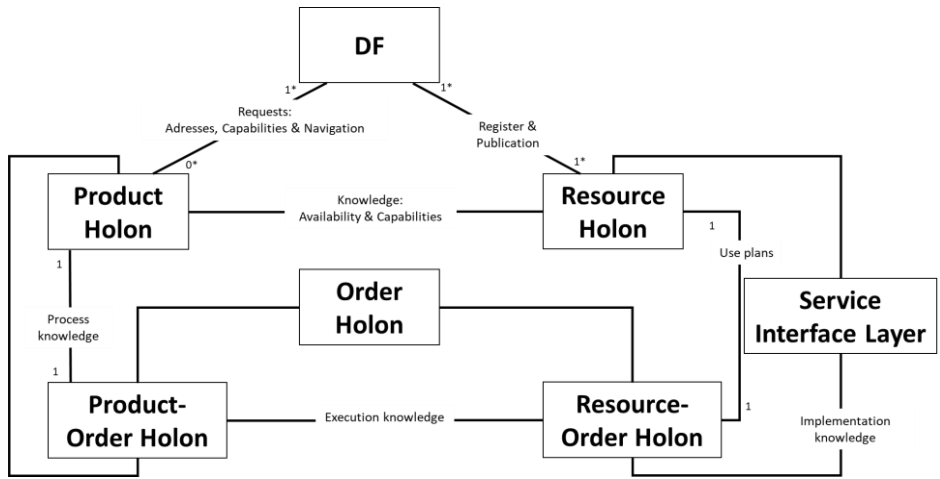

**Figure 12. SoHMS structure**

The combination of HMS and SoA paradigms is studied for many years, for example in (Bellifemine, Caire, & Greenwood, 2007; Jammes & Smit, 2005) or more recently (Giret, Garcia, & Botti, 2016; C. Morariu, Morariu, Borangiu, & Sallez, 2013). These studies outline that when the HMS provides flexibility at the structural level for the assignment of tasks and the associations between entities (holons) within the control architecture, the SoA aims at the process level for the decomposition and the encapsulation of processes that allows their distribution between resources. In a SoHMS, every operation to be performed on products are represented by services that can be offered by one or more resources. Thanks to the Mservices model, any operation in an application can have its own parameters that corresponds to the characteristics of the transformations it performs. This representation is well adapted to the repetitiveness that can be found in a manufacturing system, either in the specification of the production processes or in the specification of the capacities of a resource. Together with suitable ontologies, service discovery applications can for example be implemented. Figure 12 details the SoHMS structure.

**Table 5. Web-oriented HCA benefits to Industry 4.0 key enablers**

| Key enablers for Industry 4.0 manufacturing systems as a prerequisite | (Jammes et Smit, 2005) | (C. Morariu et al., 2013) | (Giret et al., 2016) | (Gamboa Quintanilla et al., 2016) | Comments |
|---|---|---|---|---|---|
| Sustainability | | | | | None of these architectures addresses this key enabler |
| Secure communication/ Cyber-Resilience | | | | | None of these architectures addresses this key enabler |
| Real-Time Capabilities | | | | | None of these architectures addresses this key enabler |
| Process Virtualization | | | | | None of these architectures addresses this key enabler |
| Service Orientation | X | X | X | X | This is the flagship characteristic of service-oriented HCAs |
| Interoperability | X | X | X | X | The service-orientation eases the syntactic interoperability. However, semantic interoperability is not addressed. |
| Adaptability | | | | | None of these architectures addresses this key enabler |
| Big Data Analysis | | | | | None of these architectures addresses this key enabler |
| Autonomous and decentralized Decision Support Systems | | X | X | X | The earliest works were mainly dedicated to the connectivity issues, the DSS intervened in a second step |
| Connectivity | X | X | X | X | This is the flagship ability of service-oriented HCAs |

The SoHMS architecture provides also a clear decoupling between holons that are dedicated to the decision-making process, and can potentially be located in some cloud platform, and the other ones that are dedicated to real-time control and monitoring of products and resources and are therefore implemented as close as possible to the shop-floor. This architecture premises what could be the first step towards a standardized cloud manufacturing compliant HCA. These HCAs remain at a rather high level of modeling and focus on the adaptation to the implementation platform. The objective is then to integrate the results coming from other HCAs to bridge the gap in terms of adaptability, sustainability and virtualization. Cyber-security of such architectures remains a critical issue that is not tackled yet (Wang & Haghighi, 2016). This field of research is not directly addressed, but can benefit from the recent results happening in the cyber-physical systems community (Dibaji et al., 2019). Table 5 list the different contributions of the works on web-oriented HCAs regarding the different I4.0 key enablers.

## 4.6 Digital-Twin based Architectures: towards virtualization

The concept of Digital Twin systems in Manufacturing has been taking more and more importance in the last few years. Its utilization is often seen as a standard interface between heterogeneous devices and IT infrastructure in more or less large manufacturing systems. Indeed, the possibilities it offers to induce an omniscient view of the actual state and behavior of the system is a key to many applications, from control to predictive maintenance, virtual reality or online simulation. This in-depth interoperability was integrated inside a new reference architecture, aiming to both clearly exhibit the existence of the Digital Twin and provide a new terminology, disconnected from the manufacturing one that could be misunderstood and under-appreciated in other contexts. This architecture is denoted as ARTI (Activity Resource Type Instance)(Paul Valckenaers, 2018) and represented as a cube (see figure 13).

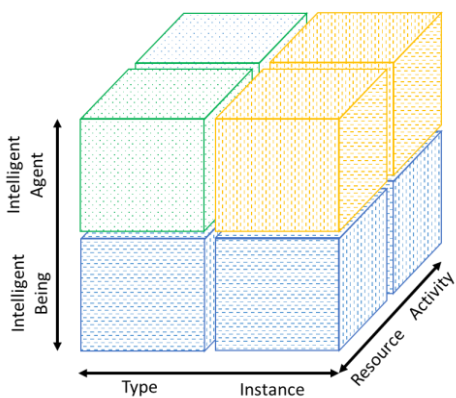

**Figure 13. ARTI reference architecture** (Paul Valckenaers, 2018)

The Digital Twin is exhibited in the blue cubes (filled with horizontal hatched lines), where every element of the physical twin can find its virtual counterpart. The yellow cubes (vertical hatched lines) and green cubes (dotted lines) are represented in this figure as control cubes, but any other software can conveniently connect to the blue cubes in order to gather the image of the current state of the physical twin. Note that blue cubes are not necessarily represented as holons, so the HCA is fully embedded inside the green and yellow cubes. This HCA is the first one exhibiting clearly a Digital Twin layer. However, the other HCAs generally embed this layer (either in a centralized or in a distributed way) as the virtualization of the physical twin is needed for a correct decision making. One major evolution of the future of HCAs is probably to cope with this emerging notion of Digital Twin in order to better integrate the whole cyber part of cyber-physical production systems.

## 4.7 Synthesis of the contributions of HCAs to the Industry4.0 key enablers requirements in manufacturing

Figure 14 introduces the evolution of the HCA with time and highlights the different trends that are currently encountered and their respective impacts on Industry 4.0. What can be highlighted is that, along time, the evolution of HCAs followed the requirements of Industry 4.0 manufacturing systems. The time taken to evolve from one requirement to another is related either to the complexity of the concepts (for Dynamic HCAs for example) or to the technological aspects (Web-oriented HCAs for example). Twenty years after the first definitions, the definition of dynamic HCAs still remains an important aspect of literature. Therefore, one might expect to find in the next few years some new HCAs targeting the last Industry 4.0 requirements that are not handled yet (even if many studies can be found evaluating how existing architectures can cope with such objectives), namely Sustainable HCAs or Cyber-Secure HCAs. These possible future trends are introduced and detailed in the next section.

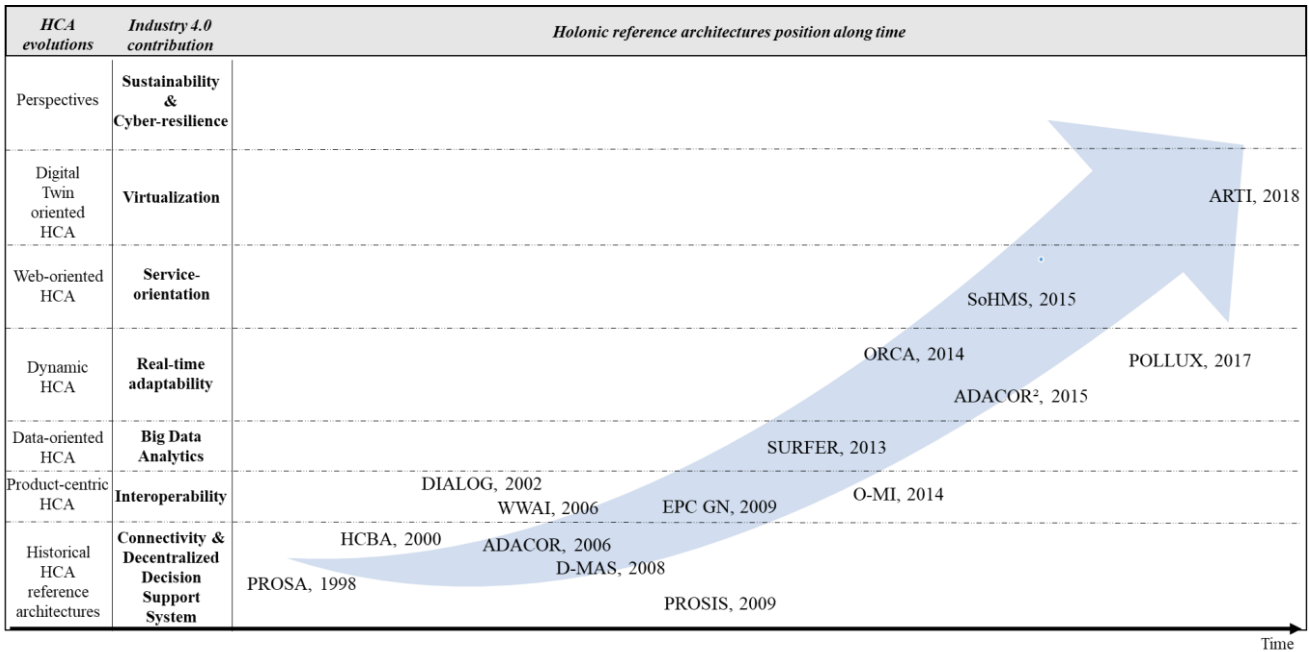

**Figure 14.** Evolution of HCAs along time and their respective contributions to Industry 4.0 manufacturing key enablers

## 5 Future trends

During these 20 years of research from the very first one, HCAs have evolved, incorporating new features all along time needed to implement (voluntary or not ) Industry 4.0 key enablers in manufacturing, as shown in Figure 14. Nevertheless, some additional challenges, offered by Industry 4.0 to researchers working on HCAs, are to be addressed. The present section introduces first the main challenges related to the key enablers identified in the previous sections and then the perspectives considering the last key enablers that are not yet addressed enough by HCAs, which could suggest opportunities to develop new researches in the field of HCAs.

*Big Data:* One important aspect of the industry 4.0 is the massive usage of sensors, monitoring the production processes. Such hyperconnected systems will generate rapidly a huge amount of unorganized data not always nor useful nor reliable. As underlined by (Babiceanu & Seker, 2016), manufacturing companies will progressively all be confronted to these data management issues grouped under the term *Big Data*. As stated before, *Learning* is a fundamental property of a holon, necessary to adapt the goals of the holon. Some attempts to couple Big Data techniques and HCAs have been proposed. As explained in previous sections, (Le Mortellec et al., 2013) defined a HCA aiming at limiting the data overload phenomenon (i.e an amount of data larger than the operator can handle) occurring with hyperconnected systems. However, far from being inappropriate, this new source of data can help to define new information such as machine health, traceability, production quality (Lee, Ardakani, Yang, & Bagheri, 2015; Lee, Kao, & Yang, 2014). Artificial Intelligence is then one key technologies (Jardim-Goncalves, Romero, & Grilo, 2017; László Monostori, 2014) to treat this sources of data. Moreover, new information of knowledge obtained via these techniques should then be integrated in the different elements of the HCAs. There have been attempts to apply Big Data techniques to manufacturing data (Lade, Ghosh, & Srinivasan, 2017; O. Morariu, Morariu, Borangiu, & Răileanu, 2018). However, some additional works are still required to integrate these techniques into HCAs, so that holons can autonomously learn from their environment. Data processing features should then be added to holonic architectures, to transform data into relevant information and/or knowledge for holons. One application of such techniques could be to help solving the myopia problem. Indeed, local decisional entities are subject to, at least, social and temporal myopia, i.e. suffer from a lack of information concerning their environment and consequences of their actions. This can lead to sub-optimal solutions or even deadlocks (Zambrano Rey et al., 2013). However, myopia is a needed characteristic of holonic architectures, ensuring a fast reaction. Big Data strategies can help to adjust this level of myopia depending on the information generated by analysing the shop-floor data. To keep the previous example related to potential fields, parameters of the potential field algorithm could be adjusted thanks to information coming from the manufacturing shop-floor.

*Adaptability – Emerging behaviour:* in the Industry 4.0 vision, systems should be able to react to unforeseen events, and to propose new behaviours, leading to potential issues relevant to unforeseen emerging behaviours. Current HCAs incorporate reaction mechanisms, based on local decisional entities, sometimes coordinated by global decisional entities. Researchers in this field that have designed HCAs had to address such an issue and develop tools to control, limit or foster emerging behaviours. Contributions are based on the control of the level of autonomy of the local holons against the level of authority of the global one. For example, in the Pollux HCA, a switching mechanism controls the degree of autonomy of holons depending on their potential future behaviours regarding expected performances and detected perturbations. Such control is never easy to achieve and can impact deeply the stability of the manufacturing system. Researchers working in the field of Industry 4.0 could gain from studying what has been proposed by researchers developing HCAs. On the contrary, what has rarely been addressed by researchers working on HCAs is the increasing diversity of possible autonomous systems evolving in Industry 4.0 systems connected using IoT technologies for example (products, robots, AGV, tools, human operators, customers, suppliers, etc.). This complexity may foster researchers working in the field of HCAs to develop more generic and open architectures, able to cope with the increasing diversity, where finally everything is holon, from the smallest intelligent sensor to the entire supply chain. The ARTI reference architecture can be seen as a first step towards this evolution.

*Adaptability – Performance guarantee:* In this same field of reaction to unforeseen events, the adaptability of the control architecture expected in Industry 4.0 vision leads to a question of performance evaluation of the systems. When a disturbance occurs, hybrid HCAs such as POLLUX, ADACOR or ORCA are meant to modify their own organisation to minimize the impact of this disturbance. A benchmarking study of HCAs is then a crucial challenge to explore. However, evaluating the performance of HCAs is reputed to be a difficult task, as it requires a dynamic evaluation of the control system's response to predefined scenario (Cardin & L'anton, 2017; Trentesaux et al., 2013). For example, (Jovanović, Zupan, Starbek, & Prebil, 2014) studied the implementation of a HCA on a "green"-tyre-manufacturing system. The objective is to evaluate how the holonic control is able to eliminate the impact of machine breakdowns on productivity. To do so, two examples of scenario are chosen, and the comparison with so called classical control approaches exhibits a 4% increase of productivity. Already difficult on an ad hoc example, this approach comes to be very difficult when trying to make it generic. Three major problematics arise:

1. How to build a generic emulation system that enables the acceleration of time for benchmarking? The inner structure of those emulation models should be made generic in order to standardize and ease their design, which appears to be quite difficult at the time being;

2. What is the best way to integrate the dynamic part of the scenarios in the architecture? As the time is speed up, emulation and control architecture needs to be synchronized. As a matter of fact, the triggering of time-related event is made difficult. Also, the triggering of event based on the behaviour of the system or of the control makes the task still more difficult as it requires the software responsible for the scenario management needs to be omniscient at full speed;

3. How to define a full set of scenarios that are representative of the diversity of systems encountered and objectives expected? A small number is already defined (Cardin & L'anton, 2017), but the number is too limited and they do not cover the whole scope of what could be expected.

To guarantee the performance of holonic architectures, formal proofs are another interesting research possibility. This possibility is also interesting with respect to ethical considerations, considering that proofs of behaviour are the first steps towards the determination of responsibilities in case of incident. To do so, three major steps have to be investigated:

- **Generic models definition**: The existing HCAs are based on models where specific and generic contents are poorly decoupled. It is therefore necessary to define generic models of both the structure and the behaviour of the HCA to enable the verification step;

- **Model-driven engineering**: This task is dedicated to proving the validity of the models by generating a fully functional HCA conform to the previous models. This task shall be supported by some expertise in model transformations;

- **Formal verification**: this task is finally meant to evaluate the feasibility of applying verification to the models defined previously in order to prove some properties of the HCA that are valuable for the production control. Considering the expected proofs, feedback loops on the models might be necessary in order to refine the models in order to fit the target.

*Service-orientation – Cloud Manufacturing*: The enhancement of connectivity on the shop-floor led to the expectations of managing the production through the network, and even through the internet. This management enables to synchronize several industrial sites during the whole chain of value creation, and even connect to suppliers, subcontractors or customers in a secure way in order to manage the workload and delivery delays in real-time. Denoted Cloud Manufacturing, this objective showed up to be much more difficult as the response time of such networks happened to be incompatible with those needed for industrial control. However, with decentralized control, such as the one introduced via HCAs, a multi-layer control architecture is built that make it possible to aggregate the information so that real-time control and security remains local whereas longer horizon decision-making is getting in the cloud. The scalability of HCAs makes it possible to integrate in the same control architecture various functions with different time scales, such as device control or supply chain synchronization. However, the reference architectures currently available only demonstrate this feasibility in a theoretical way. An innovative architecture, clearly integrating those multi-level applications and implementing natively services for multi-actor negotiations would be of a great importance in the realization of some Cloud Manufacturing generic architectures.

*Virtualization – Realizing the Digital Twin:* The concept of Digital Twin is gaining a lot of interest in the past few years in both industrial and academic communities. The objective is to be able to build a virtual replica of a manufacturing system in order to be able to predict, simulate or analyse the behaviour of the system in real-time. This twin is not yet realized, and there is not even a consensual definition that emerges from literature. The features of aggregation and autonomy inherent to holons and HCAs are of great interest for the design of the twin. Indeed, the structure of the digital twin is also meant to mimic the structure of the physical twin. This model of structure is already present in every HCA, as the control needs a representation of the current state of the system to make its own decisions. As a matter of fact, a conclusion of this observation is that a basic element constituting the core of the Digital Twin is already present in any HCA, but in a coupled way. The new HCA that will be defined need to integrate this decomposition between the decision part and the representation part, in order to exhibit the concept of Digital Twin and make it available for other applications (virtual reality for example).

*Cyber-resilience:* In essence, HCAs draw much of their effectiveness from their interconnectivity. In addition to impairing the efficiency of facility maintenance by making them more complex and interdependent, this interconnectivity poses enormous security challenges. All the developments require access to the most open network possible, whereas so far these systems have been limited to restricted, closed and proprietary ecosystems (Akella, Tang, & McMillin, 2010). Among the main threats envisaged, the injection of erroneous data and the intrusion into the network are particularly mentioned (Liu et al., 2015). (Zhang, Wang, & Tian, 2013) proposes a classification of the security threats according to the application layer to which it attacks: the sensors and actuators are vulnerable in the Perception-Execution layer, the leaks and modifications of data in the layer Transport and intrusions and loss of data privacy in the Application-Control layer. This classification exhibits the various aspects cyber-attacks can take on Industry 4.0 compliant control systems. Currently, HCAs handle the problem by securing the communication protocols, for example by encrypting the exchanges between holons. However, the distribution of the decision mechanisms among many independent holons might enable to enhance some security measures based on behaviors. These potentialities have not been integrated yet in the reference architectures, leaving the problem open for the implementation phase. Providing basic functionalities for cyber-security in the generic models of the architectures is a critical feature that should be handled in order to enhance the credibility of HCA industrial implementation.

*Sustainability – Energy efficiency*: Issues of sustainability are one of the pillars of Industry 4.0. HCAs can be very beneficial for many of them. For example, energy management through the notion of smart grid requires a massively distributed, scalable and coordinated control for which a holonic perspective will be relevant. Control of the activity of production or logistics systems will also be affected by the availability of energy from renewable energy sources. Researchers working on HCAs have recently integrated in their work such aspects and as a consequence, there is still no sound and mature contributions that could help researchers working in the field of Industry 4.0. For example, developments have been made at a general level, focusing only on general specifications to comply with. These types of constraints could benefit from a dynamic HCA in order to adapt dynamically to frequent changes in the environment (Trentesaux, Giret, Tonelli, & Skobelev, 2016). A final direction is based on the connectivity of elements consuming energy, which will give real-time access to their instantly and planned energy consumptions need for real-time decision-making. From our perspective, this aspect could constitute a starting point for both communities to work together.

*Sustainability – Integration of the human in the loop*: the human operator is a key component in Industry 4.0 (Longo, Nicoletti, & Padovano, 2017). Meanwhile, the rise of the decisional and informational complexities due to the integration of new technologies based on smart components, sensors and autonomous digital systems in future industrial system will generate strong issues regarding the successful integration of the human in decisional and control loops. "Successful" must be understood in the sense that for example the human, in order to take accurate decisions, must be aware of the situation and, in a dual way, the industrial system must assume that the human is not perfect. Classical design approaches often assume that a magic human will always be in charge, will always take good decision within fixed time spans (Trentesaux & Millot, 2016). In HCAs, humans are often treated as human holons, aside artificial holons with whom they can cooperate and collaborate. As a consequence, a pair-to-pair interaction is often suggested, which ensures such an efficient and effective integration of the human in the HCAs. Indeed, when designing HCAs, it is assumed that 1) not all every data is available (because handled by holons that behaves in a myopic way, systematically) thus deciding in an ill-structured decision context is normal, 2) autonomy of holons and openness of the HCAs facilitate the management of unexpected situations, including the support and the management of original approaches and solutions imagined by a human, and 3) it is usual and easy in the design of the HCAs to associate a companion holon to a human "holon" to facilitate his comprehension of the situation faced, thus increasing his awareness of the situation (Paul Valckenaers et al., 2011). On the other side, Industry 4.0 generates new interesting fields of research that researchers in HCA could address regarding human aspects. For example, the emerging of autonomous systems in Industry 4.0 will require to work on ethical aspects, acceptability and trust (Trentesaux & Rault, 2017). Ensuring that a holon behaves in a ethical way when interacting with human or ensuring that the human can trust holons are two of the urgent research topics. Another aspect concern the convergence of the digital and the human worlds, where concepts like symbiosis, hyper-human, augmented-human emerge (Romero, Bernus, Noran, Stahre, & Berglund, 2016).

## 6  Conclusions

Industry 4.0 is a major shift in manufacturing systems, considered as a strategic initiative worldwide. From a study of major references, 10 relevant key enablers have been identified for implementing Industry 4.0-compliant manufacturing systems and synthetized in table 1. In a second part, a description of the holonic paradigm and holon properties has been provided. It helped to identify that the holonic paradigm can theoretically answer many prerequisites of Industry 4.0. It could be seen that, if not a perfect match, HCAs may answer to many requirements of the Industry 4.0 and therefore could play a major role in the future manufacturing systems. The main reference architectures have been grouped into categories (Historical, Dynamic, Data-oriented, Product-centric, Web-oriented and Digital-Twin-based) and detailed. Each category can address specific features of the Industry 4.0. However, some key enablers of Industry 4.0 are still not addressed by HCAs or still need more research efforts. It will then be necessary either to deepen or dig into new research themes. The last part of this article underlines several research directions that could lead to the following new HCAs:

- **Data-oriented HCAs** integrating AI functions either to prevent or to exploit the Big Data. Distributed learning mechanisms could be integrated in actual HCAs in order to increase reactivity and adaptability;
- **Performance-proven Dynamic HCAs** capable to exhibit emerging behaviours while always guaranteeing a certain quality of service;
- **Digital Twin-based HCAs** that will integrate Digital Twins with the HCAs, to better monitor and plan the evolution of the controlled manufacturing system;
- **Cyber-Secure cloud-based HCAs**, implementing native services for multi-actor negotiations with secure exchanges;
- **Sustainable HCAs**, able to better control the energy consumption of manufacturing systems while fully integrating the human in the loop.

Investigating these research directions will bring answers to build sustainable, secure, totally integrated and evolutionary holonic architectures that would exhibit the different key enablers needed for a full application of Industry 4.0.

## 7 Acknowledgements

The authors would like to thank the reviewers for their valuable time and constructive comments that allowed to improve this research paper.